\documentclass[prl,aps,twocolumn,showpacs]{revtex4}

\def\btau{\mbox{\boldmath$\tau$}}
\usepackage{graphicx}
\begin{document}
\title
{Pattern formation of microtubules and motors: inelastic interaction of polar rods}
\author{Igor S. Aranson$^1$ and Lev S. Tsimring$^2$}
\affiliation{$^1$Argonne National Laboratory, 9700 South Cass
Avenue, Argonne, Illinois, 60439\\
$^2$Institute for Nonlinear Science, University of California, San
Diego, La Jolla, CA 92093-0402}
\date{\today}

\begin{abstract}
We derive a model describing spatio-temporal organization of an
array of microtubules interacting via molecular motors. Starting
from a stochastic model of inelastic polar rods with a generic
anisotropic interaction kernel we obtain a set of  equations for
the local rods concentration and orientation. At large enough mean
density of rods and concentration of motors, the model describes
orientational instability. We demonstrate that the orientational
instability leads to the formation of vortices and (for large
density and/or kernel anisotropy) asters seen in recent
experiments.
\end{abstract}

\maketitle

One of the most important functions of molecular motors (MM) is to
organize a network of long filaments (microtubules, MT) during
cell division to form cytosceletons  of daughter cells
\cite{howard00}. A number  {\em in vitro} experiments were
performed
\cite{takiguchi91,urrutia91,nedelec97,surrey01,humpgrey02,nedelec01}
to study interaction of MM and MT   in isolation from other
biophysical processes simultaneously occurring {\em in vivo}. At
large enough concentration of MM and MT, the latter organize in
{\em asters} and {\em vortices} depending on the type and
concentration of MM.

After MM binds to a microtubule at a random position, it marches
along it in a fixed direction until it unbinds without appreciable
displacement of MT.
If a MM binds to {\em two} MTs, it can change their mutual
position and orientation significantly. In Ref. \cite{surrey01},
the interaction of rod-like filaments via motor binding and motion
has been studied, and patterns resembling experimental ones were
observed. In \cite{lee01} a phenomenological model for the MM
density  and the MT orientation has been proposed.
Ref. \cite{kim03} generalized this model by including separate
densities of free and bound MM, as well as the density of MT. They
found the transition from asters to vortices as the density of MM
is increased,  in disagreement with experimental evidence
\cite{nedelec01} that the asters give way to vortices with {\em
decreasing} the MM concentration.
A   phenomenological flux-force relation for active gels was
applied  in \cite{kruse}.  While vortex and aster solutions were
obtained, an analysis  of that model is difficult because of a
large number of unknown parameters. In Ref. \cite{marchetti03}
a set of equations for MT density and orientation was  derived
by averaging conservation laws for MT
probability distribution function. However, this model does not
exhibit orientation transition for the homogeneous MT
distributions.

Here we derive  a model for the collective spatio-temporal
dynamics of MTs  starting with a master equation for interacting
inelastic polar rods. Our  model differs from the transport
equations \cite{marchetti03} in that it maintains the detailed
balance of rods with a certain orientation. The model exhibits
an onset of
orientational order for large enough density of MT and MM, formation of
vortices and then asters with increase in the MM concentration.

MMs enter the model implicitly by specifying the interaction rules
between two rods. Since the diffusion of  MMs is about 100 times
higher than that of  MTs,  at the first step we neglect spatial
variations of the MM density. While variable MM concentration
affects certain quantitative aspects \cite{nedelec01}, our
analysis captures salient features of the phenomena.
All rods are assumed to be of length $l$ and diameter $d\ll l$,
and are characterized   by the its center of mass ${\bf r}$ and
orientation  angle $\phi$.

\begin{figure}[ptb]
\includegraphics[width=3.in,angle=0]{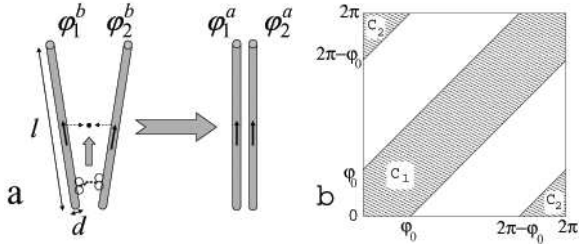}
\caption{a - sketch of motor-mediated two-rod interaction for
$\gamma=1/2$, b - integration regions $C_{1,2}$ for
Eq.(\protect\ref{master}).} \label{fig:sketch}
\end{figure}

{\em Maxwell model}. Consider the orientational dynamics only and
ignore the spatial coordinates of interacting rods (an analog of
the Maxwell model of binary collisions in kinetic theory of gases,
see e.g. \cite{bennaim00}). Since motor residence time on MT
(about 10 secs) is much smaller than the characteristic time of pattern formation
(about 1 hour),  we model MM-MT  inelastic interaction by an instantaneous
collision in which two rods change their orientations:
\begin{equation}
\left(\begin{array}{c} \phi^a_1 \\ \phi^a_2
\end{array}\right) = \left( \begin{array}{cc} \gamma   & 1-\gamma  \\
1-\gamma &  \gamma \end{array} \right) \left(\begin{array}{c} \phi^b_1
\\ \phi^b_2
\end{array}\right)
\label{collis}
\end{equation}
where $\phi_{1,2}^b$ are orientations  before and $\phi_{1,2}^a$
after the collision, and  $\gamma$ characterizes inelasticity of
collisions, and  $|\phi_2^b-\phi_1^b|<\phi_0<\pi$. The angle
between two rods is reduced after the collision by a factor
$2\gamma-1$. $\gamma=0$ corresponds to a totally elastic collision
(the rods exchange their angles, this case does not have
counterpart in MM-MT interaction context) and $\gamma=1/2$ corresponds
to a totally inelastic collision: rods acquire identical
orientation $\phi^a_{1,2}=(\phi^b_1+\phi^b_2)/2$ (see Fig.
\ref{fig:sketch},a). Here we assume that two rods only interact if
the angle between them is less than $\phi_0$. Because of
$2\pi$-periodicity, we have to add the rule of collision between
two rods with $2\pi-\phi_0<|\phi^b_2-\phi^b_1|<2\pi$.  In this
case we have to replace $\phi^{b,a}_1\to\phi^{b,a}_1+\pi,
\phi^{b,a}_2\to\phi^{b,a}_2-\pi$ in Eq. (\ref{collis}). In the
following we will only consider the case of totally inelastic rods
($\gamma=1/2$) and $\phi_0=\pi$, the generalization for arbitrary
$\gamma$ and $\phi_0$ is straightforward \cite{supplementary}. The
probability  $P(\phi)$ obeys the following master equation
%

\begin{eqnarray}
&&\partial_t P(\phi)=
D_r\partial^2_\phi P(\phi)
 +g \int_{C_1} d\phi_1d\phi_2P(\phi_1)P(\phi_2) \label{master}\\
 &&\times[\delta(\phi-
\phi_1 /2 -\phi_2/2  )
 -\delta(\phi-\phi_2)]+g \int_{C_2} d\phi_1d\phi_2
 \nonumber \\
 &&\times P(\phi_1)P(\phi_2)
 [\delta(\phi-\phi_1 / 2 - \phi_2/2 -\pi)-\delta(\phi-\phi_2)]
\nonumber
\end{eqnarray}
where  $g$ is the ``collision rate'' proportional to the
number of MM, the diffusion term $\propto D_r$ describes the thermal
fluctuations of rod orientation, and the integration domains $C_1,
C_2$ are shown in Fig.\ref{fig:sketch}a.  Changing variables $t
\to D_r t$, $P \to  g P/D_r $, $ w=\phi_2-\phi_1$, one obtains
\begin{eqnarray}
&&\partial_t P(\phi)
=\partial^2_\phi P(\phi) + \int_{-\pi}^{\pi}dw
\nonumber \\
&&\times \left[P(\phi+w/2)P(\phi-w/2) -P(\phi)P(\phi-w)\right]
\label{master2c}
\end{eqnarray}
The rescaled number density
 $\rho=
\int_0^{2\pi}P(\phi,t)d\phi$ now is proportional to {\it density
of rods multiplied by the density of motors}. Let us consider the
Fourier harmonics:
\begin{equation}
P_k=\langle e^{-ik\phi}\rangle=\frac{1}{2 \pi} \int _0^{2 \pi} d
\phi e^{-i k \phi} P(\phi,t) \label{cumm}
\end{equation}
The zeroth harmonic $P_0=\rho/2\pi=const$, and the real and
imaginary parts of $P_1$ represent the components $\tau_x= \langle
\cos \phi\rangle, \tau_y=\langle \sin \phi\rangle$ of the average
orientation vector $\btau$, $\tau_x+i\tau_y=P_1^*$.  Substituting
 (\ref{cumm}) into Eq.(\ref{master2c}) yields:
\begin{equation}
\dot P_k +( k^2+\rho)P_k=2\pi \sum_{m} P_{k-m} P_m S[\pi k/2- m
\pi] \label{Pk1a}
\end{equation}
(here $S(x)=\sin x/x$).
Due to the angular diffusion term, the magnitudes of harmonics decay
exponentially with $|k|$.
Assuming   $P_k=0$  for  $|k|>2$ one obtains   from
Eq.(\ref{Pk1a})
\begin{eqnarray}
\dot P_1 &+& P_1= P_0 P_1 2 (4 -\pi) -{8\over 3}  P_2 P_1^*
\label{Pk2} \\
 \dot P_2 &+&4P_2= -P_0 P_2 2\pi  + 2 \pi
P_1^2 \label{Pk3}
\end{eqnarray}

\begin{figure}[ptb]
\includegraphics[width=3.in,angle=-90]{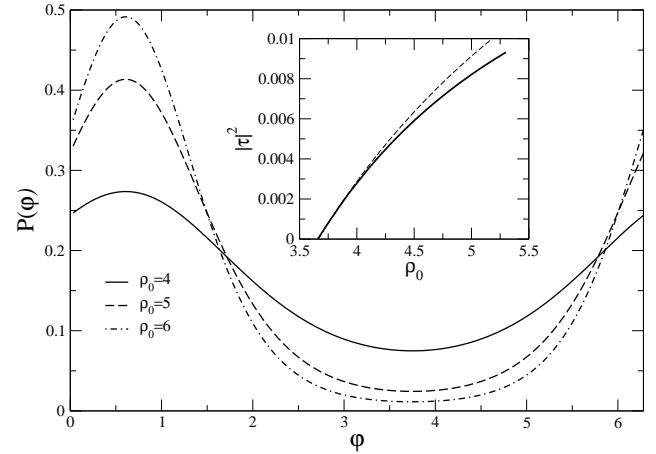}
\caption{ Stationary solutions $P(\phi)$ for different $\rho$.
Inset: the stationary value of $|\btau|$ vs  $\rho$  obtained from
the Maxwell model (\protect\ref{master2c}), dashed line -
truncated model (\protect\ref{Pk2c}).} \label{fig:distr}
\end{figure}

Since near the instability threshold the decay rate of $P_2$ is much
larger than the growth rate of $P_1$, we can neglect
the time derivative $\dot P_2$ and obtain $P_2=AP_1^2$
with $A=2\pi(\rho+4 )^{-1}$ and arrive at:
\begin{equation} \dot{\btau} =\epsilon \btau -A_0|\btau|^2\btau
\label{Pk2c}
\end{equation}
with $\epsilon=\rho  (4\pi^{-1}-1)-1\approx 0.273 \rho -1 $ and
$A_0= 8 A/3$. For large enough $\rho
>\rho_{c}=\pi/(4-\pi)\approx 3.662$, an ordering instability
leads to spontaneous rods alignment. This
instability saturates at the value determined by $\rho$. Close to
the threshold $A_0 \approx 2.18$.  Fig. \ref{fig:distr} shows
stationary solutions $P(\phi)$ obtained  from Eq.
(\ref{master2c}). As seen from the Inset, the corresponding values
of $|\btau|$ are consistent with the truncated model (\ref{Pk2c})
up to $ \rho <5.5$.

To describe the {\em  spatial localization} of interactions
we  introduce the probability distribution $P(\bf r, \phi, t)$ to
find a rod with orientation $\phi$ at location ${\bf r}$ at time
$t$. The master equation for $P({\bf r}, \phi, t)$ can be written
as
\begin{widetext}
\begin{eqnarray}
&&\partial_t P({\bf r},\phi)=
\partial^2_\phi P({\bf r},\phi) + \partial_i D_{ij}
\partial_j P({\bf r}, \phi)
 + \int\int d{\bf r_1}d{\bf r}_2\int_{-\phi_0}^{\phi_0}dw \left[
W({\bf r}_1,{\bf r}_2,\phi+w/2,\phi-w/2) \right.
\nonumber\\
&& \times P({\bf r}_1,\phi+w/2)P({\bf
r}_2,\phi-w/2)\delta\left({{\bf r}_1+{\bf r}_2\over 2}-{\bf
r}\right)  \left.- W({\bf r}_1,{\bf r}_2,\phi,\phi-w)P({\bf
r_2},\phi)P({\bf r}_1,\phi-w)\delta\left({\bf r}_2 -{\bf
r}\right)\right] \label{master3}
\end{eqnarray}
\end{widetext}
where we performed the same rescaling as in Eq.(\ref{master2c})
and dropped argument $t$ for brevity. The first two terms in
the r.h.s. of (\ref{master3}) describe angular and translational
diffusion of rods with the diffusion tensor
$D_{ij}=\frac{1}{D_r} \left(D_\parallel n_i n_j + D_\perp
(\delta_{ij}-n_i n_j)\right)$.
Here ${\bf n}=(\cos(\phi), \sin(\phi))$. $D_r, D_\parallel,
D_\perp$ are known in polymer physics:
$D_\parallel = \frac {k_B T}{\xi_\parallel}, D_\perp = \frac {k_B
T}{\xi_\perp}, D_r=\frac {4 k_B T}{\xi_r}$
where $\xi_\parallel, \xi_\perp, \xi_r $ are corresponding drag
coefficients. For rod-like molecules, $ \xi_\parallel= 2 \pi
\eta_s l /\log (l/d) ; \xi_\perp= 2 \xi_\parallel; \xi_r \approx
\pi \eta_s l^3/ 3 \log (l/d) $ where $\eta_s$ is shear viscosity
\cite{doi}.

\begin{figure}[ptb]
\includegraphics[width=3.in,angle=-90]{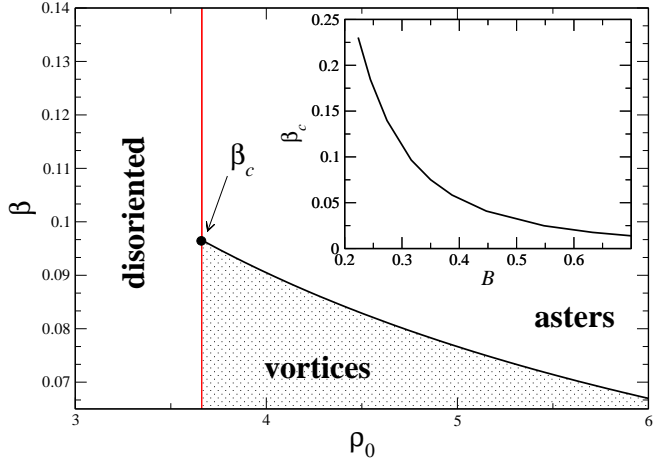}
\caption{Phase boundaries obtained form the linear stability
analysis of aster solution  for $B^2=0.05$, dashed line shows
bundling instability limit $\rho_0>\rho_b=5$. Inset: Position of
critical point $H_c$ vs $B$ at $\rho_0=4.5$. } \label{fig:bound}
\end{figure}

The last term  of Eq.(\ref{master3}) describes MM-mediated
interaction of rods. We assume that after the interaction, the two
rods  acquire the same orientation and the same spatial location
in the middle of their original locations. The interaction kernel
$W$ is localized in space, but in general does not have to be
isotropic.  On the symmetry grounds we assume the following  form
(we assume 2D geometry and neglect higher-order anisotropic
corrections): \vspace{0.1in}
\[ W= {1\over
b^2 \pi}\exp\left[-{({\bf r}_1-{\bf r}_2)^2\over b^2}\right]
(1+\beta({\bf r}_1-{\bf r}_2)\cdot({\bf n}_1-{\bf n}_2))
\]
with $b \approx l =const $. This form implies that only nearby MTs
interact effectively due to MMs. The $O(\beta)$ anisotropic term
describes the dependence of the coupling strength on the MT mutual
orientation: ``diverging'' polar rods (such as shown in Fig.
\ref{fig:sketch},a) interact stronger than ``converging'' ones.
This is the simplest term  yielding non-trivial coupling between
density and orientation. We  perform Fourier expansion in $\phi$
and truncate the series at $|n|>2$, $2 \pi P_0$ gives the local
number density $\rho({\bf r}, t)$, and $P_{\pm1}$ the local
orientation $\btau({\bf r},t)$. Omitting   calculations (see
\cite{supplementary}), rescaling  space by $l$, and introducing
dimensionless parameters $ B=b/l,H=\beta l B^2 $, we arrive at
\begin{widetext}
\begin{eqnarray}
\partial_t \rho &=&   \nabla^2 \left[ \frac{\rho}{32}
-{B^2 \rho^2 \over 16 }\right]  +{\pi B^2H\over 16} \left[ 3
\nabla \cdot  \left( \btau  \nabla^2\rho - \rho \nabla^2 \btau
\right)+
 2 \partial_i \left(
\partial_j \rho \partial_j \tau_i   - \partial_i \rho \partial_j
\tau_j \right) \right] -\frac{7 \rho_0 B^4}{256} \nabla^4 \rho
\label{rho_1}
\\
\partial_t{\btau} &=&
\frac{5}{192} \nabla^2 \btau + \frac{1}{96}\nabla (\nabla \cdot
\tau ) +\epsilon\btau -A_0|\btau|^2\btau
 + H\left[\frac{\nabla\rho^2}{16 \pi}- \left(\pi-{8\over
3}\right) \btau(\nabla\cdot\btau)- {8\over 3} (\btau \nabla) \btau
\right] +\frac{B^2 \rho_0}{4 \pi}\nabla^2 \btau   \label{tau_1}
\end{eqnarray}
\end{widetext}
The last two terms in Eqs. (\ref{rho_1}),(\ref{tau_1}) are
linearized near the mean density $\rho_0=\langle \rho \rangle$.
The last term in Eq. (\ref{rho_1}) regularizes the short-wave
instability when the diffusion term changes sign for
$\rho_0>\rho_b=1/4 B^2$. This instability leads to strong density
variations associated with formation of MT bundles, see Fig.
\ref{fig:bound}.

\begin{figure}[ptb]
\includegraphics[width=3.in,angle=-90]{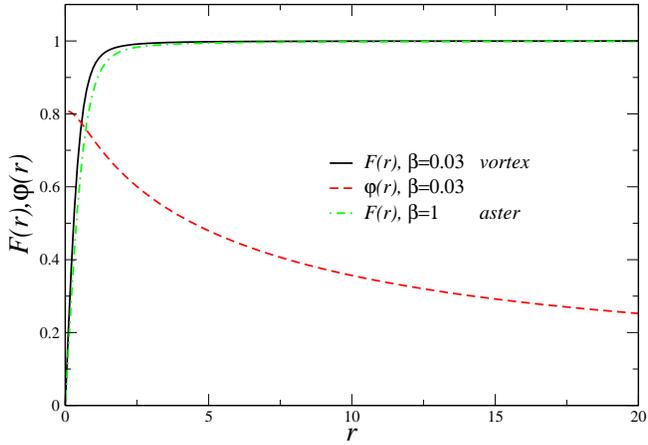}
\caption{Stationary  vortex and aster solutions $\tau_x+i
\tau_y=F(r) \exp[i \theta+ i \varphi(r)]$ to Eq. (\ref{rad2}), for
$\rho_0=4, B^2=0.05$.} \label{fig:vort_ast}
\end{figure}

{\it Aster and vortex solutions}. If $B^2H\ll 1$, the
density modulations are rather small,  and Eq. (\ref{tau_1}) for
orientation $\btau$ decouples from Eq. (\ref{rho_1}). It
is convenient to rewrite Eq. (\ref{tau_1}) for complex variable
$\psi=\tau_x+i \tau_y$ in polar coordinates $r, \theta$: $ \psi=
F(r) \exp [ i \theta + i \varphi(r)]$ where the amplitude
$F(r)$ and the phase $\varphi(r)$ are real functions.  For the aster
solution $\varphi(r)=0$ and for the vortex $\varphi(r)\ne 0$. Asters
and vortices can be examined in the framework of one-dimensional
problem for $V= \sqrt{A_0} F(r) \exp[ i \varphi(r)]$:
\begin{eqnarray}
&&\partial_t V = D_1 \Delta_r V + D_2 \Delta_r V^* + \left( 1-| V|
^2 \right) V \nonumber
\\
&&- H \left ( a_1 V \mbox{Re} \nabla_r V + a_2
\partial_r V  \mbox{Re} V+ \frac{ a_2 V \mbox{Im} V}{r} \right)
\label{rad2}\end{eqnarray} where $\Delta_r=\partial_r^2+r^{-1}
\partial_r-r^{-2}$, $\nabla_r=\partial_r+r^{-1}$, $D_1=1/32+ \rho_0 B^2/4
\pi$, $D_2=1/192$, $a_1=(\pi-8/3)/\sqrt{A_0}\approx 0.321, a_2=8
/3 \sqrt{A_0}\approx 1.81 $, and we rescaled time $t \to
t/\epsilon$ and space by $r \to r/\sqrt{\epsilon}$. The aster and
vortex solutions for certain parameter values obtained by numerical
integration of Eq. (\ref{rad2}) are shown in Fig.
\ref{fig:vort_ast}. Vortices are observed only for small values of
$H$ and give way to asters for larger $H$. For $H=0$,
Eq.(\ref{rad2}) reduces to a form which was studied in
\cite{at03}. It was shown in \cite{at03} that the term $\Delta_r V^*$
favors vortex solution $(\varphi=\pi/2)$. In
contrast, the terms proportional to $H$ select asters. Increasing
$H$ leads to gradual reduction of $\varphi$, and at a finite $H_0(\rho_0)$
$\phi(r)=0$, i.e. the
transition from vortices to asters occurs. For $0<H<H_0$, the vortex
solution has a non-trivial structure. As seen in Fig.
\ref{fig:vort_ast}, the phase $\varphi \to 0$ for $r \to \infty$,
i.e.  vortices and asters become indistinguishable far away from
the core.

The phase diagram   is shown in Fig. \ref{fig:bound}. The solid
line $H_0(\rho_0)$ separating vortices from asters is obtained from solution of
the linearized Eq. (\ref{rad2}) by tracking the most unstable
eigenvalue $\lambda$ of the aster. For this purpose the solution
to Eq. (\ref{rad2}) was sought in the form $V=F+i w \exp(\lambda
t)$, where real $w$ obeys $\hat L=\lambda w$ with operator
\begin{equation}
\hat L\equiv \bar D \Delta_r  + \left( 1-F ^2 - a_1 H  \nabla_r
F\right)  -a_2 H F \nabla_r \label{rad3}\end{equation}
($\bar D=D_1-D_2$).  Eq. (\ref{rad3})  was solved by the
matching-shooting method. The dashed line corresponds to the
orientation transition limit $\rho_0=\rho_c$. The lines meet in a
critical point $H_c=H_0(\rho_c)$ above which vortices are unstable
for arbitrary small $\epsilon>0$. The phase diagram is consistent
with experiments, see  Ref. \cite{surrey01}: for low value of
kernel anisotropy $H<H_c$ (possibly corresponding to kinesin
motors) increase of the density $\rho_0$  first leads to formation
of vortices and then asters. For  $H>H_c$ (possibly corresponding
to Ncd) only asters are observed.

For $H \ne 0$ well-separated vortices and asters exhibit
exponentially weak interaction. For asters it follows from the
fact that $\hat L$ is not a self-adjoint operator. Null-space of
$\hat L^\dagger$ exponentially decays at large $r$,  $w \sim
\exp[-r/L_0]$ with the screening length in the original length
units $L_0 = \bar D /a_2H\sqrt{\epsilon}$
 (see
\cite{ArKr}).

\begin{figure}[ptb]
\includegraphics[width=3.in,angle=0]{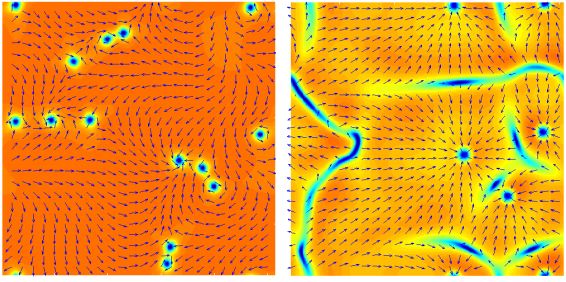}
\caption{Orientation  $\tau$ for vortices ($H=0.006$, left) and
asters ($H=0.125$, right) obtained from Eqs.
(\ref{rho_1},\ref{tau_1}). Color code indicates the intensity of
$|\tau|$ (red corresponds to maximum and blue to zero), $B^2=0.05,
\rho_0=4$, domain of integration $80\times 80$ units, time of
integration 1000 units. } \label{fig:av}
\end{figure}

We studied the full system (\ref{rho_1}),(\ref{tau_1})
numerically. Integration was performed in a two-dimensional square
domain with periodic boundary conditions by quasi-spectral method.
For small $H$ we observed vortices and for larger $H$ asters, in
agreement with the above analysis.
As seen in Fig. \ref{fig:av},
asters have unique orientation of the microtubules (here, towards
the center). Asters with opposite orientation of $\tau$ are
unstable. In large domains asters form a disordered network of
cells with a cell size of the order of $L_0$. Neighboring cells
are separated by the ``shock lines'' containing saddle-type
defects.
Starting from a random initial condition we
observed initial merging and annihilation of asters. Eventually,
annihilation slows down due to exponential weakening of the
interaction of asters.

We derived continuous equations for the evolution of MT
concentration and orientation.
We found that initially disordered  system exhibits an ordering
instability similar to a nematic phase transition in ordinary
polymers at high density. The important difference is that here
the ordering instability is mediated by MM and can occur at
arbitrary low densities of MT. At the nonlinear stage, the
instability leads to experimentally observed formation of asters
and vortices.
We thank Leo Kadanoff, Jacques Prost, and Valerii Vinokur for
useful discussions. This work was supported by the U.S. DOE,
grants W-31-109-ENG-38 (IA) and DE-FG02-04ER46135 (LT).

\newpage

\end{document}